\documentclass[
 reprint,
 superscriptaddress,
 amsmath,amssymb,
 aps,
]{revtex4-2}

\usepackage{booktabs}
\usepackage{graphicx}
\usepackage[version=4]{mhchem}
\usepackage{microtype}
\usepackage{physics}

\usepackage[hidelinks]{hyperref}

\renewcommand{\epsilon}{\varepsilon}
\renewcommand{\phi}{\varphi}

\renewcommand{\vec}[1]{\boldsymbol{\mathbf{#1}}}

\begin{document}

\title{Neural network enhanced measurement efficiency for molecular groundstates}

\author{Dmitri Iouchtchenko}
\affiliation{Department of Physics and Astronomy, University of Waterloo, Waterloo, Ontario, N2L 3G1, Canada}
\affiliation{Zapata Computing Canada Inc., 325 Front St W, Toronto, Ontario, M5V 2Y1, Canada}
\author{Jérôme F. Gonthier}
\affiliation{Zapata Computing Inc., 100 Federal Street, Boston, MA 02110, USA}
\author{Alejandro Perdomo-Ortiz}
\affiliation{Zapata Computing Canada Inc., 325 Front St W, Toronto, Ontario, M5V 2Y1, Canada}
\author{Roger G. Melko}
\affiliation{Department of Physics and Astronomy, University of Waterloo, Waterloo, Ontario, N2L 3G1, Canada}
\affiliation{Perimeter Institute for Theoretical Physics, Waterloo, Ontario, N2L 2Y5, Canada}

\date{\today}

\begin{abstract}
It is believed that one of the first useful applications for a quantum computer will be the preparation of groundstates of molecular Hamiltonians.
A crucial task involving state preparation and readout is obtaining physical observables of such states, which are typically estimated using projective measurements on the qubits.
At present, measurement data is costly and time-consuming to obtain on any quantum computing architecture, which has significant consequences for the statistical errors of estimators.
In this paper, we adapt common neural network models (restricted Boltzmann machines and recurrent neural networks) to learn complex groundstate wavefunctions for several prototypical molecular qubit Hamiltonians from typical measurement data.
By relating the accuracy $\epsilon$ of the reconstructed groundstate energy to the number of measurements, we find that using a neural network model provides a robust improvement over using single-copy measurement outcomes alone to reconstruct observables.
This enhancement yields an asymptotic scaling near $\epsilon^{-1}$ for the model-based approaches, as opposed to $\epsilon^{-2}$ in the case of classical shadow tomography.
\end{abstract}

\maketitle

\section{Introduction}

Finding the groundstate of a molecular Hamiltonian on a classical computer is a difficult task, but quantum phase estimation on a fault-tolerant quantum computer is known to efficiently obtain the groundstate energy under reasonable conditions~\cite{kitaev1995quantum,abrams1999quantum}.
However, fault tolerance remains out of reach; to date, only noisy intermediate-scale quantum (NISQ) devices have been created, where the deep circuits required for phase estimation cannot be realized.
Despite their shortcomings, it is believed that NISQ devices can still be used to assist classical ones in the task of calculating molecular groundstate energies~\cite{kandala2017hardwareefficient}.
Current approaches involve transforming the Hamiltonian from a second-quantized fermionic representation into qubit form~\cite{tranter2018comparison}, the groundstate of which could then be prepared on a quantum computer using strategies such as the variational quantum eigensolver (VQE) or other hybrid quantum--classical algorithms~\cite{mcclean2016theory}.
Once the groundstate -- or a suitable approximation -- is prepared, expectation values of physical observables can be estimated by performing measurements on the qubits.

State preparation and measurement times may vary depending on the specific experimental implementation.
Regardless, given the significant hardware limitations in present-day quantum devices, it is crucial to extract information about physical observables as efficiently as possible using the available resources,  including the collected measurement outcomes.
The gold standard for probing an experimental state is quantum state tomography, which aims to accurately reconstruct the full wavefunction or density matrix, allowing for the general extraction of the energy and other properties.
However, full state tomography comes at the cost of exponential growth of classical resources with the number of qubits, which limits its applicability to only the smallest of systems.

Beyond exact tomography, there are several promising protocols for the efficient analysis of measurement data provided by a quantum computer.
In particular, it is possible to efficiently compute the expectation value of typical observables directly from samples, using single-qubit factorable positive operator valued measures (POVMs)~\cite{carrasquilla2019reconstructing} without a model-based tomographic reconstruction, or using classical shadow tomography~\cite{huang2020predicting,hadfield2020measurements}.
In the case of typical observables like the energy, the sample complexity (defined as the number of copies $S$ of the quantum state necessary to achieve an additive error $\epsilon$) scales as $\epsilon^{-2}$ for classical shadows~\cite{huang2020predicting}.

One can also explore the accuracy of estimator reconstruction when measurement outcomes are used to train a generative model, instead of computing the expectation value directly from the samples.
Motivated by techniques in machine learning, models based on neural networks can serve as powerful wavefunction ansatzes that are readily trainable from data~\cite{carrasquilla2019reconstructing,torlai2018neuralnetwork,melko2019restricted,hibatallah2020recurrent}.
The goal of training a generative model is to learn the salient elements of the quantum state, while hopefully requiring fewer resources than state tomography when scaling to larger numbers of qubits~\cite{sehayek2019learnability}.
Indeed, it was shown in Ref.~\cite{torlai2020precise} that training a restricted Boltzmann machine (RBM)~\cite{torlai2016learning,carleo2017solving} and then sampling it (the ``generative'' step) significantly reduced the variance of the energy estimator of some simple molecular Hamiltonians, at the possible cost of a bias due to training imperfections.

In this paper, we implement two different generative models as wavefunction anstatzes: complex RBMs~\cite{carleo2017solving,torlai2018neuralnetwork} and complex recurrent neural networks (RNNs)~\cite{hibatallah2020recurrent}.
We use these models to reconstruct the groundstates of standard molecular Hamiltonians for \ce{LiH}, \ce{BeH2}, and \ce{H2} from synthetic single-copy measurement data.
We perform a comparison of the errors in the energy and -- when possible -- the fidelity, between the models, maximum likelihood pure state tomography, and classical shadows.
We find that for both the RBM and RNN, the sample complexity scales approximately as $\epsilon^{-1}$, as compared to $\epsilon^{-2}$ for classical shadows.
Our results suggest that generative models may offer a robust advantage in reconstructing physical observables in cases where data is limited, as in present-day quantum computing devices.

\section{Models}

In this section, we provide details of the generative models that we use to reconstruct the molecular groundstates from qubit measurement data.
We assume that, for a given basis, each projective measurement outcome is a vector with $N$ binary values (one per qubit), and that these outcomes are distributed according to the Born rule.
However, unlike standard generative models used for classical machine learning tasks, we require that our models are capable of fully describing a pure quantum state $\ket{\psi}$, whose coefficients have both an amplitude and a phase.
We employ two of the most well-studied generative models suited to this purpose: the complex restricted Boltzmann machine (RBM) and the complex recurrent neural network (RNN).
The complex RBM takes on the form described in Ref.~\cite{carleo2017solving}: a regular RBM, but with complex-valued parameters.
In the standard basis, the complex RBM state has the coefficients
\begin{subequations}
\begin{align}
	\ip{\vec{\sigma}}{\phi_\mathrm{RBM}(\vec{\lambda})}
	&= \frac{1}{\sqrt{Z(\vec{\lambda})}} \sum_{\vec{h}} e^{
			\vec{h}^\mathrm{T} \vec{W} \vec{\sigma}
			+ \vec{b} \cdot \vec{\sigma}
			+ \vec{c} \cdot \vec{h}
		} \\
	&= \frac{1}{\sqrt{Z(\vec{\lambda})}} e^{\vec{b} \cdot \vec{\sigma}}
		\prod_{j=1}^{N_\mathrm{h}}
			\left[ 1 + e^{(\vec{W} \vec{\sigma} + \vec{c})_j} \right],
\end{align}
where
\begin{align}
	Z(\vec{\lambda})
	&= \sum_{\vec{\sigma}} \abs{
			\sum_{\vec{h}} e^{
				\vec{h}^\mathrm{T} \vec{W} \vec{\sigma}
				+ \vec{b} \cdot \vec{\sigma}
				+ \vec{c} \cdot \vec{h}
			}
		}^2
\end{align}
\end{subequations}
is the squared normalization, $\vec{\lambda}$ represents the parameters ($\vec{W}$, $\vec{b}$, $\vec{c}$), and $\vec{h}$ is the length-$N_\mathrm{h}$ hidden state vector.

The complex RNN is the same as in Ref.~\cite{hibatallah2020recurrent}, but we omit the softsign function for the phases.
It has the standard-basis coefficients
\begin{subequations}
\begin{align}
	\ip{\vec{\sigma}}{\phi_\mathrm{RNN}(\vec{\lambda})}
	&= \prod_{j=1}^N e^{2\pi i \theta_j(\vec{\sigma})} \sqrt{p_j(\vec{\sigma})},
\end{align}
where
\begin{align}
	\theta_j(\vec{\sigma})
	&= (\vec{u} \cdot \vec{h}_j(\vec{\sigma}) + a) \sigma_j
			+ \vec{v} \cdot \vec{h}_j(\vec{\sigma}) + b
\end{align}
is the local phase,
\begin{align}
	p_j(\vec{\sigma})
	&= \frac{e^{(\vec{w} \cdot \vec{h}_j(\vec{\sigma}) + c) \sigma_j}}{1 + e^{\vec{w} \cdot \vec{h}_j(\vec{\sigma}) + c}}
\end{align}
is the conditional probability,
\begin{align}
	\vec{h}_j(\vec{\sigma})
	&= \tanh(\vec{M} \vec{h}_{j-1}(\vec{\sigma}) + \vec{p} \sigma_{j-1} + \vec{q})
	\label{eq:hidden-state}
\end{align}
\end{subequations}
is the length-$N_\mathrm{h}$ hidden state (with $\tanh$ as the element-wise nonlinearity), $\vec{\lambda}$ represents the parameters ($\vec{M}$, $\vec{p}$, $\vec{q}$, $\vec{u}$, $\vec{v}$, $\vec{w}$, $a$, $b$, $c$), and $N$ is the length of the basis state vector $\vec{\sigma}$.
We do not employ a one-hot encoding, and we initialize the hidden state recurrence with $\vec{h}_0 = (0, \ldots, 0)$ and $\sigma_0 = 0$.

In the typical case of reconstructing a single probability distribution, RBMs and RNNs are trained by minimizing the cross entropy term of the Kullback--Leibler divergence.
However, for a state $\ket{\psi}$ with both an amplitude and a phase, projective measurements must be performed in multiple orthonormal bases to capture sufficient information.
This results in multiple probability distributions, and a natural approach is to combine all the cross entropies into a single loss function~\cite{torlai2020precise}:
\begin{align}
	\mathcal{L}(\vec{\lambda})
	&= -\frac{1}{\abs{\mathcal{D}}} \sum_{k=1}^K
		\sum_{\vec{\sigma} \in \mathcal{D}_k}
			\log{\abs{
				\mel{\vec{\sigma}}{\hat{R}_k^\dagger}{\phi(\vec{\lambda})}
			}^2},
\end{align}
where $\hat{R}_k$ is a unitary operator that maps the standard basis to measurement basis $k$, $\mathcal{D}_k$ is the collection of standard-basis measurement outcomes obtained from the rotated target state $\hat{R}_k^\dagger \ket{\psi}$, $K$ is the number of bases (or ``settings'') used for projective measurements, and $\abs{\mathcal{D}} = \sum_{k=1}^K \abs{\mathcal{D}_k}$.
Further model and training details are provided in Appendix~\ref{sec:training}.

In addition, for the small system sizes considered in this work ($N = 4$, $6$, and $8$ qubits), it is feasible to represent the full state as a length-$2^N$ vector with complex coefficients $\ip{\vec{\sigma}}{\phi_\mathrm{WF}(\vec{\lambda})}$; the parameters $\vec{\lambda}$ of this ``model'' are simply the coefficients themselves, which can be learned via a maximum likelihood method.
At a minimum of the loss function $\mathcal{L}(\vec{\lambda})$, the corresponding state $\ket{\phi(\vec{\lambda})}$ satisfies
\begin{align}
	\ket{\phi(\vec{\lambda})}
	&= \frac{1}{\abs{\mathcal{D}}} \sum_{k=1}^K
		\sum_{\vec{\sigma} \in \mathcal{D}_k}
			\frac{
				\hat{R}_k \ket{\vec{\sigma}}
			}{
				\mel{\phi(\vec{\lambda})}{\hat{R}_k}{\vec{\sigma}}
			}
	= T(\ket{\phi(\vec{\lambda})}).
\end{align}
The nonlinear operator $T$ resembles the composite physical imposition operator of Ref.~\cite{goyeneche2014quantum}, albeit with squared overlap magnitudes, measurement data in place of exact outcome probabilities, and summation instead of composition over the measurement settings.
We exploit this similarity to find the coefficients of the complex state vector at a critical point of the loss function by starting from the exact state $\ket{\psi}$ and applying $T$ until a fixed point is reached:
\begin{subequations}
\begin{align}
	\ket{\phi_\mathrm{WF}(\vec{\lambda})}
	&= \tilde{T}(\tilde{T}(\cdots \tilde{T}(\tilde{T}(\ket{\psi})) \cdots )),
	\label{eq:fixed-point}
\end{align}
where
\begin{align}
	\tilde{T}(\ket{\phi})
	&= \frac{
			\ket{\phi} + T(\ket{\phi})
		}{
			\norm{\ket{\phi} + T(\ket{\phi})}
		}
\end{align}
\end{subequations}
includes a contribution from the previous state to smooth out oscillations near convergence.
This wavefunction optimization process yields a reference state with which to compare the other data-driven reconstructions using the generative models.

\section{Results and discussion}

We are interested in reconstructing observables for groundstates of three molecular Hamiltonians $\hat{H}$ for \ce{LiH}, \ce{BeH2}, and \ce{H2}.
In order to implement these in a form suitable for a quantum computer, each fermionic Hamiltonian is decomposed into a qubit Hamiltonian as a sum of Pauli strings.
This is obtained via the parity transformation implemented in Qiskit~\cite{qiskit}; see Appendix~\ref{sec:hamiltonians} for more details.

Each of the qubit Hamiltonians is sufficiently small that its groundstate $\ket{\psi}$ can be found using exact diagonalization.
This allows us to calculate the exact energies (or any generic observable), as well as to produce a synthetic measurement dataset of arbitrary size.
For each synthetic measurement outcome from a given molecular groundstate, an orthonormal basis $k$ is chosen uniformly at random from the $K$ bases determined by the Pauli string expansion of the corresponding Hamiltonian, and an outcome label $\vec{\sigma}$ is selected with the Born rule probability
\begin{align}
	p_k(\vec{\sigma})
	&= \abs{\mel{\vec{\sigma}}{\hat{R}^\dagger_k}{\psi}}^2.
\end{align}
The resulting datasets are used to train the RBM and RNN generative models.
By way of comparison, we use the same datasets to implement maximum likelihood tomography on all $2 \times 2^N$ real-valued parameters of the full wavefunction via fixed point iteration.
We also compare to uniform classical shadows by implementing Algorithm~1 of Ref.~\cite{hadfield2020measurements}.

\begin{figure}
	\includegraphics{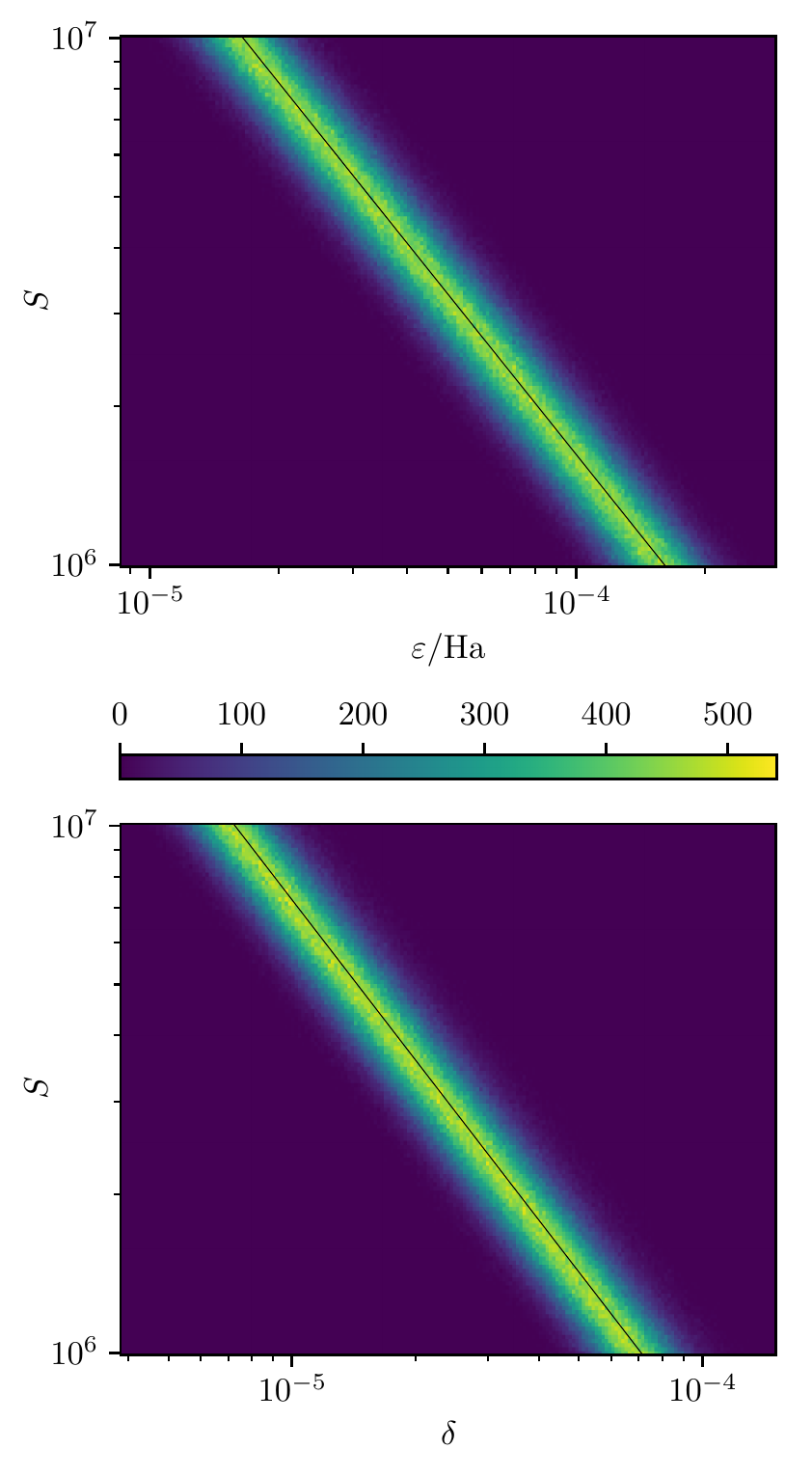}
	\caption{
		Distributions of $\epsilon$ (top panel) and $\delta$ (bottom panel) for wavefunction optimization of the groundstate of \ce{BeH2} using the fixed point iteration in Eq.~\ref{eq:fixed-point}.
		The colors indicate the number of calculations that had access to $S$ measurement outcomes and produced a state with the corresponding $\epsilon$ or $\delta$.
		The lines are fits from Fig.~\ref{fig:scaling-LiH-BeH2}.
	}
	\label{fig:histogram-BeH2}
\end{figure}

For a given groundstate, the sample complexity $S$ is defined as the number of measurements (i.e.\@ the number of copies of the state) required to achieve a target accuracy of the reconstructed state or observable.
This formulation implies that $S$ is a function of the accuracy parameters $\epsilon$ and $\delta$.
However, in practice, one must fix $S$ and obtain a state with some energy and fidelity, which will be random variables that depend on both the observed measurement outcomes and the details of the reconstruction method.
Figure~\ref{fig:histogram-BeH2} reveals that the distributions of
\begin{subequations}
\begin{align}
	\epsilon
	&= \mel{\phi(\vec{\lambda})}{\hat{H}}{\phi(\vec{\lambda})}
			- \mel{\psi}{\hat{H}}{\psi}
\end{align}
and
\begin{align}
	\delta
	&= 1 - \abs{\ip{\phi(\vec{\lambda})}{\psi}}^2
\end{align}
\end{subequations}
for maximum likelihood tomography of the \ce{BeH2} groundstate vary smoothly with $S$.
Therefore, we examine averages over multiple independent datasets with fixed $S$ in order to draw conclusions about the scaling of $S$.

\begin{figure}
	\includegraphics{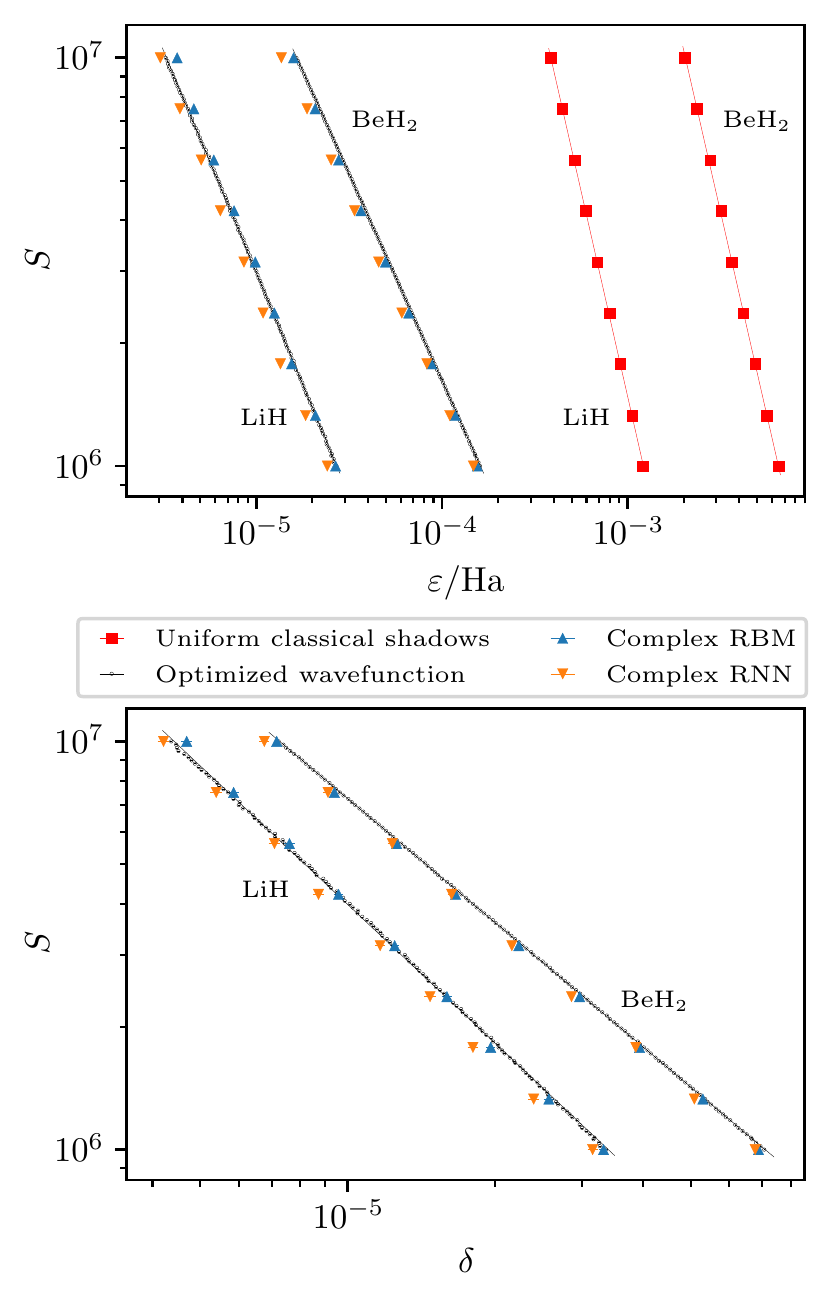}
	\caption{
		The number of samples $S$ required to reach an energy error $\epsilon$ (top panel) and infidelity $\delta$ (bottom panel) for \ce{LiH} and \ce{BeH2}.
		The markers show the means of several calculations at a fixed $S$, and the horizontal error bars show the standard errors of the mean.
		The lines are least squares fits through the markers; their slopes are given in Table~\ref{tab:slopes}.
	}
	\label{fig:scaling-LiH-BeH2}
\end{figure}

\begin{figure}
	\includegraphics{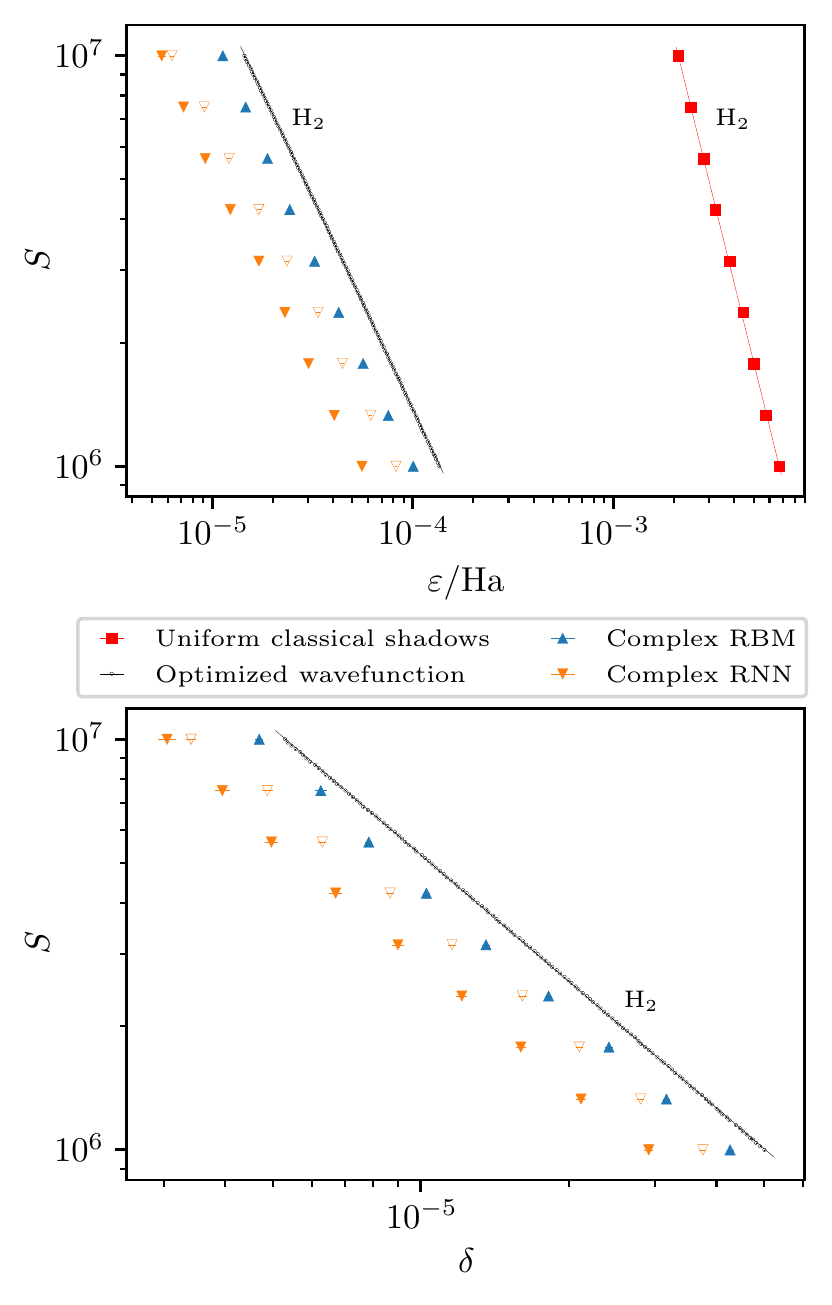}
	\caption{
		The number of samples $S$ required to reach an energy error $\epsilon$ (top panel) and infidelity $\delta$ (bottom panel) for \ce{H2}.
		The markers show the means of several calculations at a fixed $S$, and the horizontal error bars show the standard errors of the mean.
		The lines are least squares fits through the markers; their slopes are given in Table~\ref{tab:slopes}.
		Data for the hollow triangles was generated using a larger model; see the hyperparameters in Appendix~\ref{sec:training}.
	}
	\label{fig:scaling-H2}
\end{figure}

Such averages are shown for all the methods in Figs.~\ref{fig:scaling-LiH-BeH2} and \ref{fig:scaling-H2}.
For uniform classical shadows, the estimated energy is used directly to compute the energy error:
\begin{align}
	\epsilon
	&= \abs{
			\bar{E}
			- \mel{\psi}{\hat{H}}{\psi}
		}.
\end{align}
We chose to work in the regime of $10^6$ to $10^7$ measurements, as this is currently tractable on some experimental hardware.
For example, on the order of $10^7$ measurement outcomes can be obtained per hour using a superconducting quantum computer~\cite{arute2019quantum} (although for other hardware, e.g.\@ neutral atom arrays, a realistic number may be several orders of magnitude less~\cite{endres2016atombyatom}).
With this many measurement outcomes, we found that linear fits are appropriate for model-based reconstructions ($R^2 > 0.998$), optimized wavefunctions ($R^2 > 0.999$), and uniform classical shadows ($R^2 > 0.999$); their slopes are summarized in Table~\ref{tab:slopes}.
Since the data is log-transformed for plotting, the slopes of the fits are the exponents of the power law expressions relating the sample complexity $S$ to the estimates of reconstruction quality ($\epsilon$ and $\delta$).

\begin{table}
	\caption{
		Slopes of the linear fits in Figs.~\ref{fig:scaling-LiH-BeH2} and \ref{fig:scaling-H2} (for visual clarity, fits for the complex RBM and RNN are not drawn).
	}
	\label{tab:slopes}
	\begin{tabular}{c@{\hskip 1em} *{3}{c c}}
		\toprule
		& \multicolumn{2}{c}{\ce{LiH}} & \multicolumn{2}{c}{\ce{BeH2}} & \multicolumn{2}{c}{\ce{H2}} \\
		\cmidrule(lr){2-3}
		\cmidrule(lr){4-5}
		\cmidrule(lr){6-7}
		& $\epsilon$ & $\delta$ & $\epsilon$ & $\delta$ & $\epsilon$ & $\delta$ \\
		\midrule
		Shadows      & $-2.01$ & ---     & $-2.00$ & ---     & $-1.99$ & ---     \\
		Wavefunction & $-1.09$ & $-1.13$ & $-1.01$ & $-1.01$ & $-1.03$ & $-1.02$ \\
		Complex RBM  & $-1.17$ & $-1.18$ & $-1.00$ & $-1.01$ & $-1.05$ & $-1.05$ \\
		Complex RNN  & $-1.12$ & $-1.16$ & $-0.97$ & $-1.00$ & $-0.99$ & $-1.01$ \\
		\bottomrule
	\end{tabular}
\end{table}

Assuming that results in this region of $S$ are in the asymptotic limit, we may compare the scaling between the methods mentioned above.
As listed in Table~\ref{tab:slopes}, uniform classical shadows have a sample complexity of approximately $S \propto \epsilon^{-2}$, as expected from the method's structure as a direct average over random samples.
On the other hand, the model-based approaches get very near to scaling as $S \propto \epsilon^{-1}$.
This implies that -- in the small error limit -- significantly fewer measurements are needed to achieve the same quality of approximation.
In the present case of small $N$, the fidelity $1 - \delta$ between the model and target wavefunctions can be computed directly; as shown in Table~\ref{tab:slopes}, the model-based approaches are close to $S \propto \delta^{-1}$ using only single-copy measurements.

\begin{figure}
	\includegraphics{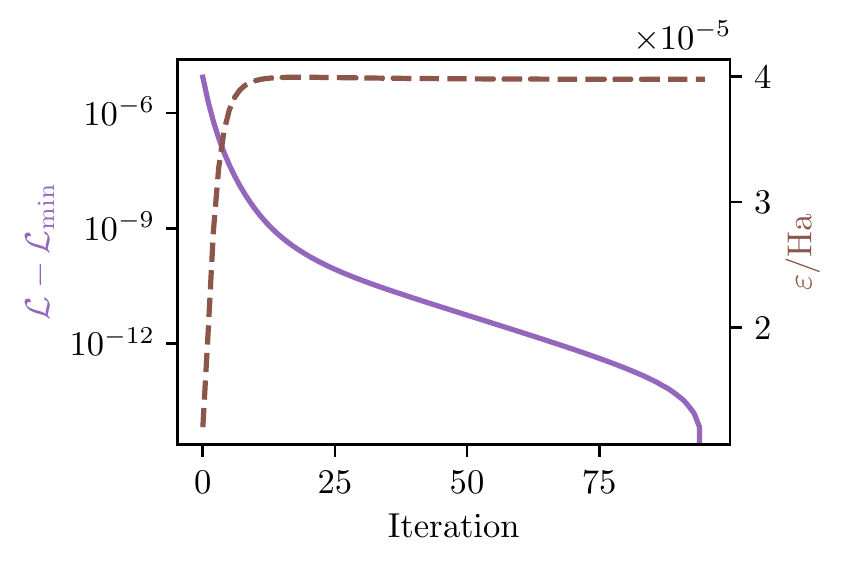}
	\caption{
		Wavefunction optimization trajectory for \ce{H2}, starting from the state of a trained complex RNN.
		The loss $\mathcal{L}$ (solid, left axis) decreases to $\mathcal{L}_\mathrm{min}$ as the optimization proceeds, but the energy error $\epsilon$ (dashed, right axis) increases more than twofold.
	}
	\label{fig:traj}
\end{figure}

One concern with models that are optimized to reproduce a finite collection of measurement outcomes is the possibility of overfitting to the training data.
Indeed, we consistently observe this with the optimized wavefunctions, the vast majority ($>99.998\%$) of which end up at a minimum of the loss function after several rounds of fixed point iteration.
Although it is not always possible to find a state in the Hilbert space that perfectly reproduces the observations, being in a loss minimum suggests reasonably good memorization of the training data.
In contrast, the states produced by the trained complex RBMs and RNNs never completely minimize the loss; states with a smaller loss always exist, but either the models cannot express them or the training procedure cannot find them.
For example, the optimization trajectory in Fig.~\ref{fig:traj} demonstrates that fixed point iteration can easily improve the loss of a trained complex RNN wavefunction (better memorization) at the cost of a larger energy error (worse generalization, as described in Appendix~\ref{sec:generalization}).

The effect of this trade-off can be observed in Fig.~\ref{fig:scaling-H2} for \ce{H2}, where the models invariably generalize better than the optimized wavefunctions.
Using larger models (which are expected to be more expressive and possibly more trainable) can also negatively impact the ability to generalize, as shown by the hollow triangles in Fig.~\ref{fig:scaling-H2}.
It is likely that a description of the actual ground state of \ce{H2} is much simpler (in the sense of Kolmogorov complexity) than a description of any particular realization of measurement data, so a more constrained model might stumble upon the former while searching for the latter.
Thus, the limited flexibility of the models can be a boon for groundstate energy estimation in practice.
Additionally, if one is willing to consider a variational Monte Carlo approach, overfitting can be avoided altogether by switching to an energy-based loss after the initial portion of the training~\cite{bennewitz2021neural,czischek2022dataenhanced}.

\section{Conclusions and outlook}

In this paper, we explore the use of generative models adapted from modern machine learning for the efficient reconstruction of observables from prototypical qubit Hamiltonians in quantum chemistry: \ce{LiH}, \ce{BeH2}, and \ce{H2}.
The neural network ansatzes we use are well-capable of representing groundstate wavefunctions, can be made systematically more expressive (by increasing e.g.\@ the number of hidden units), and have well-studied and efficient training heuristics.
Using synthetic datasets of typical noiseless experimental measurements on the groundstate wavefunction, we calculate the sample complexity $S$: the number of independent measurements required to reproduce the energy or fidelity up to a given accuracy.

We find that the energy reconstruction obtained by training a generative model results in an energy error $\epsilon$ that is orders of magnitude smaller than that achieved by classical shadow tomography using the same amount of measurement data.
More generally, these findings suggest that $S$ for the model-based reconstruction should scale asymptotically as $\epsilon^{-1}$, as opposed to $\epsilon^{-2}$ for classical shadows.

Of course, in practice, this type of improvement would depend on the model training heuristics scaling favorably.
While very few results relevant to this scaling exist in the machine learning literature, efficient scaling of the training procedure has been observed previously in reconstructions of many-qubit groundstates relevant for condensed matter physics~\cite{sehayek2019learnability}.

Our results illustrate that some applications of present-day quantum computers can benefit from the use of generative models adapted from machine learning.
For the small molecules studied here, we can conclude that a model-based approach is better able to make use of hardware with a data rate limited by state preparation and measurement than alternatives such as classical shadow tomography.
More generally, hybrid quantum--classical algorithms such as VQE can benefit from modelling quantum states using neural networks.
These approaches require multiple states to be prepared on a quantum computer, with very many measurements performed over the course of the algorithm~\cite{wecker2015progress,gonthier2020identifying}.
Insertion of neural network models into the hybrid workflow has the potential to drastically reduce the amount of quantum computation involved, enabling the study of systems which would otherwise be prohibitively expensive.

\begin{acknowledgments}
The authors would like to thank Yi Hong Teoh, Marta Mauri, Juan Carrasquilla, Alexander A. Kunitsa, Peter D. Johnson, and Jhonathan Romero for insightful discussions.
This work was supported by Mitacs through the Mitacs Elevate program.
RGM acknowledges support from the Natural Sciences and Engineering Research Council of Canada (NSERC), the Canada Research Chair (CRC) program, the New Frontiers in Research Fund, and the Perimeter Institute for Theoretical Physics.
Research at the Perimeter Institute is supported in part by the Government of Canada through the Department of Innovation, Science and Economic Development Canada and by the Province of Ontario through the Ministry of Economic Development, Job Creation and Trade.
The numerical calculations in this work were possible thanks to the computational resources provided by Compute Canada.
\end{acknowledgments}

\appendix

\section{Models and training}
\label{sec:training}

The complex RBMs and RNNs were trained using the Adam optimizer~\cite{kingma2017adam} in the Flux machine learning library for Julia~\cite{innes2018flux,innes2018fashionable}, with the hyperparameters given in Table~\ref{tab:training}.
The hyperparameters were chosen to provide a reasonable level of convergence of the energy simultaneously for small and large $S$.
Since the $S$ measurements are allocated to the $K$ bases uniformly at random, approximately $S/K$ measurements are performed in each basis.

\begin{table}
	\caption{
		Model and training hyperparameters.
		The number of real-valued parameters is $N_\mathrm{P} = 2 (N N_\mathrm{h} + N + N_\mathrm{h})$ for complex RBMs and $N_\mathrm{P} = N_\mathrm{h}^2 + 5 N_\mathrm{h} + 3$ for complex RNNs.
		Hyperparameters for the large \ce{H2} RNN ($N_\mathrm{h} = 64$, hollow triangles in Fig.~\ref{fig:scaling-H2}) are also included.
	}
	\label{tab:training}
	\begin{tabular}{rl r@{\hskip 2mm} r cc}
		\toprule
		Molecule & Model & $N_\mathrm{h}$ & $N_\mathrm{P}$ & Learning rate & Epochs \\
		\midrule
		\ce{LiH} & RBM & 3 & 38 & $3 \times 10^{-3}$ & $1.6 \times 10^5$ \\
		\ce{LiH} & RNN & 4 & 39 & $1 \times 10^{-3}$ & $1.0 \times 10^5$ \\[1mm]
		\ce{BeH2} & RBM & 9 & 138 & $3 \times 10^{-4}$ & $1.0 \times 10^5$ \\
		\ce{BeH2} & RNN & 15 & 303 & $1 \times 10^{-4}$ & $4.5 \times 10^5$ \\[1mm]
		\ce{H2} & RBM & 5 & 106 & $3 \times 10^{-3}$ & $1.2 \times 10^5$ \\
		\ce{H2} & RNN & 9 & 129 & $3 \times 10^{-4}$ & $1.2 \times 10^5$ \\
		\cmidrule{1-6}
		\ce{H2} & RNN & 64 & 4419 & $1 \times 10^{-5}$ & $1.2 \times 10^5$ \\
		\bottomrule
	\end{tabular}
\end{table}

\section{Molecular Hamiltonians}
\label{sec:hamiltonians}

We studied three molecules near their equilibrium bond lengths: \ce{LiH}, \ce{BeH2}, and \ce{H2}.
In the case of \ce{BeH2}, we only considered linear geometries with equal bond lengths.
The electronic Hamiltonians were transformed into qubit form using Qiskit's \texttt{FermionicTransformation} class with Psi4~\cite{smith2020psi4} as the driver and the mapping set to \texttt{FermionicQubitMappingType.PARITY}.
Energies are reported in hartrees (Ha).
Per-molecule information is provided in Table~\ref{tab:hamiltonians}.

\begin{table}
	\caption{
		Details of the molecular Hamiltonians.
		The transformation parameters were chosen to produce the same system sizes $N$ as in Ref.~\cite{torlai2020precise}.
	}
	\label{tab:hamiltonians}
	\begin{tabular}{r@{\hskip 1em} *{2}{c@{\hskip 1em}} c}
		\toprule
		& \ce{LiH} & \ce{BeH2} & \ce{H2} \\
		\midrule
		Bond length & 1.55~\r{A} &1.32~\r{A} & 0.75~\r{A} \\
		Basis set & STO-3G & STO-3G & 6-31G \\
		Spin orbitals & 12 & 14 & 8 \\[1mm]
		\texttt{freeze\_core} & \texttt{True} & \texttt{True} & \texttt{False} \\
		\texttt{orbital\_reduction} & \texttt{[-3, -2]} & \texttt{[-3]}  & \texttt{None} \\
		\texttt{two\_qubit\_reduction} & \texttt{True} & \texttt{True} & \texttt{False} \\
		\texttt{z2symmetry\_reduction} & \texttt{None} & \texttt{[+1, +1]} & \texttt{None} \\[1mm]
		Qubits ($N$) & 4 & 6 & 8 \\
		Pauli terms & 100 & 216 & 185 \\
		Measurement bases ($K$) & 25 & 83 & 54 \\
		\bottomrule
	\end{tabular}
\end{table}

\section{Energy error as a proxy for generalization}
\label{sec:generalization}

One might intuitively expect that a smaller additive energy error $\epsilon$ indicates that the generative model is better able to generalize to unseen configurations.
For example, if the model's energy is computed by Monte Carlo, evaluating the estimator on samples that are more representative of the true distribution should provide a result that is closer to the true energy.

\begin{figure}
	\includegraphics{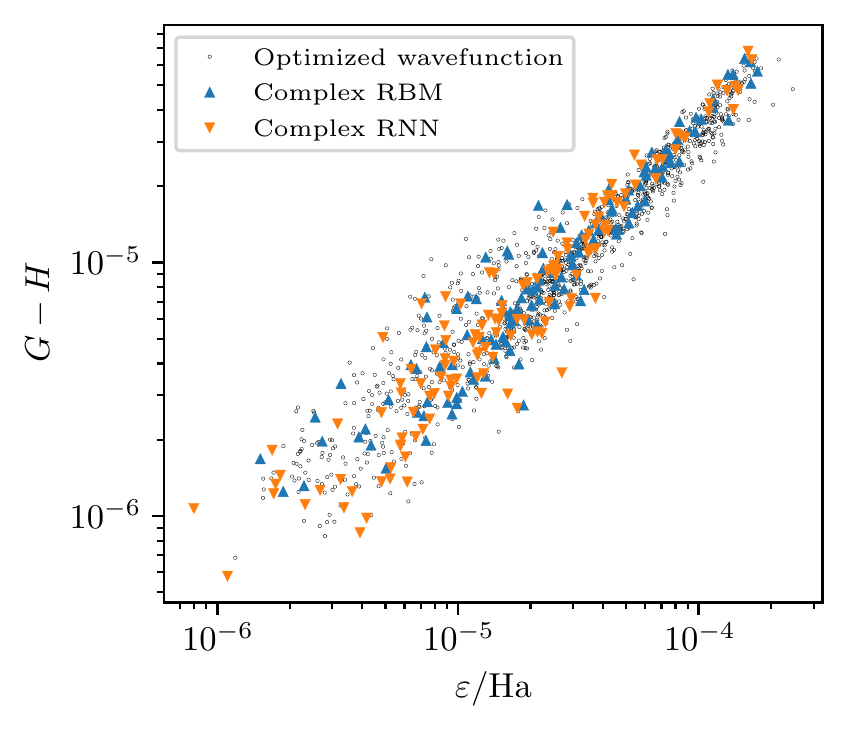}
	\caption{
		Correlation between the generalization error $G$ and energy error $\epsilon$ for all 3 molecules.
		The average entropy $H$ of the target distributions is subtracted from each $G$ in order to make the values commensurable between molecules.
	}
	\label{fig:generalization}
\end{figure}

The complex RBM and RNN are trained to reproduce a particular finite collection of data using the empirical risk (the cross entropy relative to the histogram of the data).
Hence, a suitable measure of generalization error is the true risk (the cross entropy relative to the target distribution)~\cite{hansen1996unsupervised},
\begin{align}
	G(\vec{\lambda})
	&= -\frac{1}{K} \sum_{k=1}^K
		\sum_{\vec{\sigma}}
			p_k(\vec{\sigma})
			\log{\abs{
				\mel{\vec{\sigma}}{\hat{R}_k^\dagger}{\phi(\vec{\lambda})}
			}^2},
\end{align}
which is obtained in the limit of infinite data.
As can be seen in Fig.~\ref{fig:generalization}, for the wavefunctions obtained in this work, there is a strong correlation between the generalization error $G$ and the energy error $\epsilon$, suggesting that the latter is representative of the former.

\begin{thebibliography}{28}%
\makeatletter
\providecommand \@ifxundefined [1]{%
 \@ifx{#1\undefined}
}%
\providecommand \@ifnum [1]{%
 \ifnum #1\expandafter \@firstoftwo
 \else \expandafter \@secondoftwo
 \fi
}%
\providecommand \@ifx [1]{%
 \ifx #1\expandafter \@firstoftwo
 \else \expandafter \@secondoftwo
 \fi
}%
\providecommand \natexlab [1]{#1}%
\providecommand \enquote  [1]{``#1''}%
\providecommand \bibnamefont  [1]{#1}%
\providecommand \bibfnamefont [1]{#1}%
\providecommand \citenamefont [1]{#1}%
\providecommand \href@noop [0]{\@secondoftwo}%
\providecommand \href [0]{\begingroup \@sanitize@url \@href}%
\providecommand \@href[1]{\@@startlink{#1}\@@href}%
\providecommand \@@href[1]{\endgroup#1\@@endlink}%
\providecommand \@sanitize@url [0]{\catcode `\\12\catcode `\$12\catcode
  `\&12\catcode `\#12\catcode `\^12\catcode `\_12\catcode `\%12\relax}%
\providecommand \@@startlink[1]{}%
\providecommand \@@endlink[0]{}%
\providecommand \url  [0]{\begingroup\@sanitize@url \@url }%
\providecommand \@url [1]{\endgroup\@href {#1}{\urlprefix }}%
\providecommand \urlprefix  [0]{URL }%
\providecommand \Eprint [0]{\href }%
\providecommand \doibase [0]{https://doi.org/}%
\providecommand \selectlanguage [0]{\@gobble}%
\providecommand \bibinfo  [0]{\@secondoftwo}%
\providecommand \bibfield  [0]{\@secondoftwo}%
\providecommand \translation [1]{[#1]}%
\providecommand \BibitemOpen [0]{}%
\providecommand \bibitemStop [0]{}%
\providecommand \bibitemNoStop [0]{.\EOS\space}%
\providecommand \EOS [0]{\spacefactor3000\relax}%
\providecommand \BibitemShut  [1]{\csname bibitem#1\endcsname}%
\let\auto@bib@innerbib\@empty
\bibitem [{\citenamefont {Kitaev}(1995)}]{kitaev1995quantum}%
  \BibitemOpen
  \bibfield  {author} {\bibinfo {author} {\bibfnamefont {A.~Y.}\ \bibnamefont
  {Kitaev}},\ }\bibfield  {title} {\bibinfo {title} {Quantum measurements and
  the {{Abelian}} stabilizer problem},\ }\href@noop {} {\bibfield  {journal}
  {\bibinfo  {journal} {arXiv}\ } (\bibinfo {year} {1995})},\ \Eprint
  {https://arxiv.org/abs/quant-ph/9511026} {arXiv:quant-ph/9511026}
  \BibitemShut {NoStop}%
\bibitem [{\citenamefont {Abrams}\ and\ \citenamefont
  {Lloyd}(1999)}]{abrams1999quantum}%
  \BibitemOpen
  \bibfield  {author} {\bibinfo {author} {\bibfnamefont {D.~S.}\ \bibnamefont
  {Abrams}}\ and\ \bibinfo {author} {\bibfnamefont {S.}~\bibnamefont {Lloyd}},\
  }\bibfield  {title} {\bibinfo {title} {Quantum algorithm providing
  exponential speed increase for finding eigenvalues and eigenvectors},\ }\href
  {https://doi.org/10.1103/PhysRevLett.83.5162} {\bibfield  {journal} {\bibinfo
   {journal} {Phys. Rev. Lett.}\ }\textbf {\bibinfo {volume} {83}},\ \bibinfo
  {pages} {5162} (\bibinfo {year} {1999})}\BibitemShut {NoStop}%
\bibitem [{\citenamefont {Kandala}\ \emph {et~al.}(2017)\citenamefont
  {Kandala}, \citenamefont {Mezzacapo}, \citenamefont {Temme}, \citenamefont
  {Takita}, \citenamefont {Brink}, \citenamefont {Chow},\ and\ \citenamefont
  {Gambetta}}]{kandala2017hardwareefficient}%
  \BibitemOpen
  \bibfield  {author} {\bibinfo {author} {\bibfnamefont {A.}~\bibnamefont
  {Kandala}}, \bibinfo {author} {\bibfnamefont {A.}~\bibnamefont {Mezzacapo}},
  \bibinfo {author} {\bibfnamefont {K.}~\bibnamefont {Temme}}, \bibinfo
  {author} {\bibfnamefont {M.}~\bibnamefont {Takita}}, \bibinfo {author}
  {\bibfnamefont {M.}~\bibnamefont {Brink}}, \bibinfo {author} {\bibfnamefont
  {J.~M.}\ \bibnamefont {Chow}},\ and\ \bibinfo {author} {\bibfnamefont
  {J.~M.}\ \bibnamefont {Gambetta}},\ }\bibfield  {title} {\bibinfo {title}
  {Hardware-efficient variational quantum eigensolver for small molecules and
  quantum magnets},\ }\href {https://doi.org/10.1038/nature23879} {\bibfield
  {journal} {\bibinfo  {journal} {Nature}\ }\textbf {\bibinfo {volume} {549}},\
  \bibinfo {pages} {242} (\bibinfo {year} {2017})}\BibitemShut {NoStop}%
\bibitem [{\citenamefont {Tranter}\ \emph {et~al.}(2018)\citenamefont
  {Tranter}, \citenamefont {Love}, \citenamefont {Mintert},\ and\ \citenamefont
  {Coveney}}]{tranter2018comparison}%
  \BibitemOpen
  \bibfield  {author} {\bibinfo {author} {\bibfnamefont {A.}~\bibnamefont
  {Tranter}}, \bibinfo {author} {\bibfnamefont {P.~J.}\ \bibnamefont {Love}},
  \bibinfo {author} {\bibfnamefont {F.}~\bibnamefont {Mintert}},\ and\ \bibinfo
  {author} {\bibfnamefont {P.~V.}\ \bibnamefont {Coveney}},\ }\bibfield
  {title} {\bibinfo {title} {A comparison of the
  {{Bravyi}}\textendash{{Kitaev}} and {{Jordan}}\textendash{{Wigner}}
  transformations for the quantum simulation of quantum chemistry},\ }\href
  {https://doi.org/10.1021/acs.jctc.8b00450} {\bibfield  {journal} {\bibinfo
  {journal} {J. Chem. Theory Comput.}\ }\textbf {\bibinfo {volume} {14}},\
  \bibinfo {pages} {5617} (\bibinfo {year} {2018})}\BibitemShut {NoStop}%
\bibitem [{\citenamefont {McClean}\ \emph {et~al.}(2016)\citenamefont
  {McClean}, \citenamefont {Romero}, \citenamefont {Babbush},\ and\
  \citenamefont {{Aspuru-Guzik}}}]{mcclean2016theory}%
  \BibitemOpen
  \bibfield  {author} {\bibinfo {author} {\bibfnamefont {J.~R.}\ \bibnamefont
  {McClean}}, \bibinfo {author} {\bibfnamefont {J.}~\bibnamefont {Romero}},
  \bibinfo {author} {\bibfnamefont {R.}~\bibnamefont {Babbush}},\ and\ \bibinfo
  {author} {\bibfnamefont {A.}~\bibnamefont {{Aspuru-Guzik}}},\ }\bibfield
  {title} {\bibinfo {title} {The theory of variational hybrid quantum-classical
  algorithms},\ }\href {https://doi.org/10.1088/1367-2630/18/2/023023}
  {\bibfield  {journal} {\bibinfo  {journal} {New J. Phys.}\ }\textbf {\bibinfo
  {volume} {18}},\ \bibinfo {pages} {023023} (\bibinfo {year}
  {2016})}\BibitemShut {NoStop}%
\bibitem [{\citenamefont {Carrasquilla}\ \emph {et~al.}(2019)\citenamefont
  {Carrasquilla}, \citenamefont {Torlai}, \citenamefont {Melko},\ and\
  \citenamefont {Aolita}}]{carrasquilla2019reconstructing}%
  \BibitemOpen
  \bibfield  {author} {\bibinfo {author} {\bibfnamefont {J.}~\bibnamefont
  {Carrasquilla}}, \bibinfo {author} {\bibfnamefont {G.}~\bibnamefont
  {Torlai}}, \bibinfo {author} {\bibfnamefont {R.~G.}\ \bibnamefont {Melko}},\
  and\ \bibinfo {author} {\bibfnamefont {L.}~\bibnamefont {Aolita}},\
  }\bibfield  {title} {\bibinfo {title} {Reconstructing quantum states with
  generative models},\ }\href {https://doi.org/10.1038/s42256-019-0028-1}
  {\bibfield  {journal} {\bibinfo  {journal} {Nat. Mach. Intell.}\ }\textbf
  {\bibinfo {volume} {1}},\ \bibinfo {pages} {155} (\bibinfo {year}
  {2019})}\BibitemShut {NoStop}%
\bibitem [{\citenamefont {Huang}\ \emph {et~al.}(2020)\citenamefont {Huang},
  \citenamefont {Kueng},\ and\ \citenamefont {Preskill}}]{huang2020predicting}%
  \BibitemOpen
  \bibfield  {author} {\bibinfo {author} {\bibfnamefont {H.-Y.}\ \bibnamefont
  {Huang}}, \bibinfo {author} {\bibfnamefont {R.}~\bibnamefont {Kueng}},\ and\
  \bibinfo {author} {\bibfnamefont {J.}~\bibnamefont {Preskill}},\ }\bibfield
  {title} {\bibinfo {title} {Predicting many properties of a quantum system
  from very few measurements},\ }\href
  {https://doi.org/10.1038/s41567-020-0932-7} {\bibfield  {journal} {\bibinfo
  {journal} {Nat. Phys.}\ }\textbf {\bibinfo {volume} {16}},\ \bibinfo {pages}
  {1050} (\bibinfo {year} {2020})}\BibitemShut {NoStop}%
\bibitem [{\citenamefont {Hadfield}\ \emph {et~al.}(2020)\citenamefont
  {Hadfield}, \citenamefont {Bravyi}, \citenamefont {Raymond},\ and\
  \citenamefont {Mezzacapo}}]{hadfield2020measurements}%
  \BibitemOpen
  \bibfield  {author} {\bibinfo {author} {\bibfnamefont {C.}~\bibnamefont
  {Hadfield}}, \bibinfo {author} {\bibfnamefont {S.}~\bibnamefont {Bravyi}},
  \bibinfo {author} {\bibfnamefont {R.}~\bibnamefont {Raymond}},\ and\ \bibinfo
  {author} {\bibfnamefont {A.}~\bibnamefont {Mezzacapo}},\ }\bibfield  {title}
  {\bibinfo {title} {Measurements of quantum {{Hamiltonians}} with
  locally-biased classical shadows},\ }\href@noop {} {\bibfield  {journal}
  {\bibinfo  {journal} {arXiv}\ } (\bibinfo {year} {2020})},\ \Eprint
  {https://arxiv.org/abs/2006.15788} {arXiv:2006.15788} \BibitemShut {NoStop}%
\bibitem [{\citenamefont {Torlai}\ \emph {et~al.}(2018)\citenamefont {Torlai},
  \citenamefont {Mazzola}, \citenamefont {Carrasquilla}, \citenamefont
  {Troyer}, \citenamefont {Melko},\ and\ \citenamefont
  {Carleo}}]{torlai2018neuralnetwork}%
  \BibitemOpen
  \bibfield  {author} {\bibinfo {author} {\bibfnamefont {G.}~\bibnamefont
  {Torlai}}, \bibinfo {author} {\bibfnamefont {G.}~\bibnamefont {Mazzola}},
  \bibinfo {author} {\bibfnamefont {J.}~\bibnamefont {Carrasquilla}}, \bibinfo
  {author} {\bibfnamefont {M.}~\bibnamefont {Troyer}}, \bibinfo {author}
  {\bibfnamefont {R.}~\bibnamefont {Melko}},\ and\ \bibinfo {author}
  {\bibfnamefont {G.}~\bibnamefont {Carleo}},\ }\bibfield  {title} {\bibinfo
  {title} {Neural-network quantum state tomography},\ }\href
  {https://doi.org/10.1038/s41567-018-0048-5} {\bibfield  {journal} {\bibinfo
  {journal} {Nat. Phys.}\ }\textbf {\bibinfo {volume} {14}},\ \bibinfo {pages}
  {447} (\bibinfo {year} {2018})}\BibitemShut {NoStop}%
\bibitem [{\citenamefont {Melko}\ \emph {et~al.}(2019)\citenamefont {Melko},
  \citenamefont {Carleo}, \citenamefont {Carrasquilla},\ and\ \citenamefont
  {Cirac}}]{melko2019restricted}%
  \BibitemOpen
  \bibfield  {author} {\bibinfo {author} {\bibfnamefont {R.~G.}\ \bibnamefont
  {Melko}}, \bibinfo {author} {\bibfnamefont {G.}~\bibnamefont {Carleo}},
  \bibinfo {author} {\bibfnamefont {J.}~\bibnamefont {Carrasquilla}},\ and\
  \bibinfo {author} {\bibfnamefont {J.~I.}\ \bibnamefont {Cirac}},\ }\bibfield
  {title} {\bibinfo {title} {Restricted {{Boltzmann}} machines in quantum
  physics},\ }\href {https://doi.org/10.1038/s41567-019-0545-1} {\bibfield
  {journal} {\bibinfo  {journal} {Nat. Phys.}\ }\textbf {\bibinfo {volume}
  {15}},\ \bibinfo {pages} {887} (\bibinfo {year} {2019})}\BibitemShut
  {NoStop}%
\bibitem [{\citenamefont {{Hibat-Allah}}\ \emph {et~al.}(2020)\citenamefont
  {{Hibat-Allah}}, \citenamefont {Ganahl}, \citenamefont {Hayward},
  \citenamefont {Melko},\ and\ \citenamefont
  {Carrasquilla}}]{hibatallah2020recurrent}%
  \BibitemOpen
  \bibfield  {author} {\bibinfo {author} {\bibfnamefont {M.}~\bibnamefont
  {{Hibat-Allah}}}, \bibinfo {author} {\bibfnamefont {M.}~\bibnamefont
  {Ganahl}}, \bibinfo {author} {\bibfnamefont {L.~E.}\ \bibnamefont {Hayward}},
  \bibinfo {author} {\bibfnamefont {R.~G.}\ \bibnamefont {Melko}},\ and\
  \bibinfo {author} {\bibfnamefont {J.}~\bibnamefont {Carrasquilla}},\
  }\bibfield  {title} {\bibinfo {title} {Recurrent neural network wave
  functions},\ }\href {https://doi.org/10.1103/PhysRevResearch.2.023358}
  {\bibfield  {journal} {\bibinfo  {journal} {Phys. Rev. Research}\ }\textbf
  {\bibinfo {volume} {2}},\ \bibinfo {pages} {023358} (\bibinfo {year}
  {2020})}\BibitemShut {NoStop}%
\bibitem [{\citenamefont {Sehayek}\ \emph {et~al.}(2019)\citenamefont
  {Sehayek}, \citenamefont {Golubeva}, \citenamefont {Albergo}, \citenamefont
  {Kulchytskyy}, \citenamefont {Torlai},\ and\ \citenamefont
  {Melko}}]{sehayek2019learnability}%
  \BibitemOpen
  \bibfield  {author} {\bibinfo {author} {\bibfnamefont {D.}~\bibnamefont
  {Sehayek}}, \bibinfo {author} {\bibfnamefont {A.}~\bibnamefont {Golubeva}},
  \bibinfo {author} {\bibfnamefont {M.~S.}\ \bibnamefont {Albergo}}, \bibinfo
  {author} {\bibfnamefont {B.}~\bibnamefont {Kulchytskyy}}, \bibinfo {author}
  {\bibfnamefont {G.}~\bibnamefont {Torlai}},\ and\ \bibinfo {author}
  {\bibfnamefont {R.~G.}\ \bibnamefont {Melko}},\ }\bibfield  {title} {\bibinfo
  {title} {Learnability scaling of quantum states: {{Restricted Boltzmann}}
  machines},\ }\href {https://doi.org/10.1103/PhysRevB.100.195125} {\bibfield
  {journal} {\bibinfo  {journal} {Phys. Rev. B}\ }\textbf {\bibinfo {volume}
  {100}},\ \bibinfo {pages} {195125} (\bibinfo {year} {2019})}\BibitemShut
  {NoStop}%
\bibitem [{\citenamefont {Torlai}\ \emph {et~al.}(2020)\citenamefont {Torlai},
  \citenamefont {Mazzola}, \citenamefont {Carleo},\ and\ \citenamefont
  {Mezzacapo}}]{torlai2020precise}%
  \BibitemOpen
  \bibfield  {author} {\bibinfo {author} {\bibfnamefont {G.}~\bibnamefont
  {Torlai}}, \bibinfo {author} {\bibfnamefont {G.}~\bibnamefont {Mazzola}},
  \bibinfo {author} {\bibfnamefont {G.}~\bibnamefont {Carleo}},\ and\ \bibinfo
  {author} {\bibfnamefont {A.}~\bibnamefont {Mezzacapo}},\ }\bibfield  {title}
  {\bibinfo {title} {Precise measurement of quantum observables with
  neural-network estimators},\ }\href
  {https://doi.org/10.1103/PhysRevResearch.2.022060} {\bibfield  {journal}
  {\bibinfo  {journal} {Phys. Rev. Research}\ }\textbf {\bibinfo {volume}
  {2}},\ \bibinfo {pages} {022060(R)} (\bibinfo {year} {2020})}\BibitemShut
  {NoStop}%
\bibitem [{\citenamefont {Torlai}\ and\ \citenamefont
  {Melko}(2016)}]{torlai2016learning}%
  \BibitemOpen
  \bibfield  {author} {\bibinfo {author} {\bibfnamefont {G.}~\bibnamefont
  {Torlai}}\ and\ \bibinfo {author} {\bibfnamefont {R.~G.}\ \bibnamefont
  {Melko}},\ }\bibfield  {title} {\bibinfo {title} {Learning thermodynamics
  with {{Boltzmann}} machines},\ }\href
  {https://doi.org/10.1103/PhysRevB.94.165134} {\bibfield  {journal} {\bibinfo
  {journal} {Phys. Rev. B}\ }\textbf {\bibinfo {volume} {94}},\ \bibinfo
  {pages} {165134} (\bibinfo {year} {2016})}\BibitemShut {NoStop}%
\bibitem [{\citenamefont {Carleo}\ and\ \citenamefont
  {Troyer}(2017)}]{carleo2017solving}%
  \BibitemOpen
  \bibfield  {author} {\bibinfo {author} {\bibfnamefont {G.}~\bibnamefont
  {Carleo}}\ and\ \bibinfo {author} {\bibfnamefont {M.}~\bibnamefont
  {Troyer}},\ }\bibfield  {title} {\bibinfo {title} {Solving the quantum
  many-body problem with artificial neural networks},\ }\href
  {https://doi.org/10.1126/science.aag2302} {\bibfield  {journal} {\bibinfo
  {journal} {Science}\ }\textbf {\bibinfo {volume} {355}},\ \bibinfo {pages}
  {602} (\bibinfo {year} {2017})}\BibitemShut {NoStop}%
\bibitem [{\citenamefont {Goyeneche}\ and\ \citenamefont {{de la
  Torre}}(2014)}]{goyeneche2014quantum}%
  \BibitemOpen
  \bibfield  {author} {\bibinfo {author} {\bibfnamefont {D.}~\bibnamefont
  {Goyeneche}}\ and\ \bibinfo {author} {\bibfnamefont {A.~C.}\ \bibnamefont
  {{de la Torre}}},\ }\bibfield  {title} {\bibinfo {title} {Quantum tomography
  meets dynamical systems and bifurcations theory},\ }\href
  {https://doi.org/10.1063/1.4881855} {\bibfield  {journal} {\bibinfo
  {journal} {J. Math. Phys.}\ }\textbf {\bibinfo {volume} {55}},\ \bibinfo
  {pages} {062103} (\bibinfo {year} {2014})}\BibitemShut {NoStop}%
\bibitem [{\citenamefont {ANIS}\ \emph {et~al.}(2021)\citenamefont {ANIS},
  \citenamefont {Abraham}, \citenamefont {{AduOffei}}, \citenamefont {Agarwal},
  \citenamefont {Agliardi}, \citenamefont {Aharoni}, \citenamefont {Akhalwaya},
  \citenamefont {Aleksandrowicz}, \citenamefont {Alexander}, \citenamefont
  {Amy}, \citenamefont {Anagolum}, \citenamefont {Arbel}, \citenamefont
  {Asfaw}, \citenamefont {Athalye}, \citenamefont {Avkhadiev} \emph
  {et~al.}}]{qiskit}%
  \BibitemOpen
  \bibfield  {author} {\bibinfo {author} {\bibfnamefont {M.~S.}\ \bibnamefont
  {ANIS}}, \bibinfo {author} {\bibfnamefont {H.}~\bibnamefont {Abraham}},
  \bibinfo {author} {\bibnamefont {{AduOffei}}}, \bibinfo {author}
  {\bibfnamefont {R.}~\bibnamefont {Agarwal}}, \bibinfo {author} {\bibfnamefont
  {G.}~\bibnamefont {Agliardi}}, \bibinfo {author} {\bibfnamefont
  {M.}~\bibnamefont {Aharoni}}, \bibinfo {author} {\bibfnamefont {I.~Y.}\
  \bibnamefont {Akhalwaya}}, \bibinfo {author} {\bibfnamefont {G.}~\bibnamefont
  {Aleksandrowicz}}, \bibinfo {author} {\bibfnamefont {T.}~\bibnamefont
  {Alexander}}, \bibinfo {author} {\bibfnamefont {M.}~\bibnamefont {Amy}},
  \bibinfo {author} {\bibfnamefont {S.}~\bibnamefont {Anagolum}}, \bibinfo
  {author} {\bibfnamefont {E.}~\bibnamefont {Arbel}}, \bibinfo {author}
  {\bibfnamefont {A.}~\bibnamefont {Asfaw}}, \bibinfo {author} {\bibfnamefont
  {A.}~\bibnamefont {Athalye}}, \bibinfo {author} {\bibfnamefont
  {A.}~\bibnamefont {Avkhadiev}}, \emph {et~al.},\ }\href
  {https://doi.org/10.5281/zenodo.2573505} {\bibinfo {title} {Qiskit: {{An}}
  open-source framework for quantum computing}} (\bibinfo {year}
  {2021})\BibitemShut {NoStop}%
\bibitem [{\citenamefont {Arute}\ \emph {et~al.}(2019)\citenamefont {Arute},
  \citenamefont {Arya}, \citenamefont {Babbush}, \citenamefont {Bacon},
  \citenamefont {Bardin}, \citenamefont {Barends}, \citenamefont {Biswas},
  \citenamefont {Boixo}, \citenamefont {Brandao}, \citenamefont {Buell},
  \citenamefont {Burkett}, \citenamefont {Chen}, \citenamefont {Chen},
  \citenamefont {Chiaro}, \citenamefont {Collins} \emph
  {et~al.}}]{arute2019quantum}%
  \BibitemOpen
  \bibfield  {author} {\bibinfo {author} {\bibfnamefont {F.}~\bibnamefont
  {Arute}}, \bibinfo {author} {\bibfnamefont {K.}~\bibnamefont {Arya}},
  \bibinfo {author} {\bibfnamefont {R.}~\bibnamefont {Babbush}}, \bibinfo
  {author} {\bibfnamefont {D.}~\bibnamefont {Bacon}}, \bibinfo {author}
  {\bibfnamefont {J.~C.}\ \bibnamefont {Bardin}}, \bibinfo {author}
  {\bibfnamefont {R.}~\bibnamefont {Barends}}, \bibinfo {author} {\bibfnamefont
  {R.}~\bibnamefont {Biswas}}, \bibinfo {author} {\bibfnamefont
  {S.}~\bibnamefont {Boixo}}, \bibinfo {author} {\bibfnamefont {F.~G. S.~L.}\
  \bibnamefont {Brandao}}, \bibinfo {author} {\bibfnamefont {D.~A.}\
  \bibnamefont {Buell}}, \bibinfo {author} {\bibfnamefont {B.}~\bibnamefont
  {Burkett}}, \bibinfo {author} {\bibfnamefont {Y.}~\bibnamefont {Chen}},
  \bibinfo {author} {\bibfnamefont {Z.}~\bibnamefont {Chen}}, \bibinfo {author}
  {\bibfnamefont {B.}~\bibnamefont {Chiaro}}, \bibinfo {author} {\bibfnamefont
  {R.}~\bibnamefont {Collins}}, \emph {et~al.},\ }\bibfield  {title} {\bibinfo
  {title} {Quantum supremacy using a programmable superconducting processor},\
  }\href {https://doi.org/10.1038/s41586-019-1666-5} {\bibfield  {journal}
  {\bibinfo  {journal} {Nature}\ }\textbf {\bibinfo {volume} {574}},\ \bibinfo
  {pages} {505} (\bibinfo {year} {2019})}\BibitemShut {NoStop}%
\bibitem [{\citenamefont {Endres}\ \emph {et~al.}(2016)\citenamefont {Endres},
  \citenamefont {Bernien}, \citenamefont {Keesling}, \citenamefont {Levine},
  \citenamefont {Anschuetz}, \citenamefont {Krajenbrink}, \citenamefont
  {Senko}, \citenamefont {Vuletic}, \citenamefont {Greiner},\ and\
  \citenamefont {Lukin}}]{endres2016atombyatom}%
  \BibitemOpen
  \bibfield  {author} {\bibinfo {author} {\bibfnamefont {M.}~\bibnamefont
  {Endres}}, \bibinfo {author} {\bibfnamefont {H.}~\bibnamefont {Bernien}},
  \bibinfo {author} {\bibfnamefont {A.}~\bibnamefont {Keesling}}, \bibinfo
  {author} {\bibfnamefont {H.}~\bibnamefont {Levine}}, \bibinfo {author}
  {\bibfnamefont {E.~R.}\ \bibnamefont {Anschuetz}}, \bibinfo {author}
  {\bibfnamefont {A.}~\bibnamefont {Krajenbrink}}, \bibinfo {author}
  {\bibfnamefont {C.}~\bibnamefont {Senko}}, \bibinfo {author} {\bibfnamefont
  {V.}~\bibnamefont {Vuletic}}, \bibinfo {author} {\bibfnamefont
  {M.}~\bibnamefont {Greiner}},\ and\ \bibinfo {author} {\bibfnamefont {M.~D.}\
  \bibnamefont {Lukin}},\ }\bibfield  {title} {\bibinfo {title} {Atom-by-atom
  assembly of defect-free one-dimensional cold atom arrays},\ }\href
  {https://doi.org/10.1126/science.aah3752} {\bibfield  {journal} {\bibinfo
  {journal} {Science}\ }\textbf {\bibinfo {volume} {354}},\ \bibinfo {pages}
  {1024} (\bibinfo {year} {2016})}\BibitemShut {NoStop}%
\bibitem [{\citenamefont {Bennewitz}\ \emph {et~al.}(2021)\citenamefont
  {Bennewitz}, \citenamefont {Hopfmueller}, \citenamefont {Kulchytskyy},
  \citenamefont {Carrasquilla},\ and\ \citenamefont
  {Ronagh}}]{bennewitz2021neural}%
  \BibitemOpen
  \bibfield  {author} {\bibinfo {author} {\bibfnamefont {E.~R.}\ \bibnamefont
  {Bennewitz}}, \bibinfo {author} {\bibfnamefont {F.}~\bibnamefont
  {Hopfmueller}}, \bibinfo {author} {\bibfnamefont {B.}~\bibnamefont
  {Kulchytskyy}}, \bibinfo {author} {\bibfnamefont {J.}~\bibnamefont
  {Carrasquilla}},\ and\ \bibinfo {author} {\bibfnamefont {P.}~\bibnamefont
  {Ronagh}},\ }\bibfield  {title} {\bibinfo {title} {Neural error mitigation of
  near-term quantum simulations},\ }\href@noop {} {\bibfield  {journal}
  {\bibinfo  {journal} {arXiv}\ } (\bibinfo {year} {2021})},\ \Eprint
  {https://arxiv.org/abs/2105.08086} {arXiv:2105.08086} \BibitemShut {NoStop}%
\bibitem [{\citenamefont {Czischek}\ \emph {et~al.}(2022)\citenamefont
  {Czischek}, \citenamefont {Moss}, \citenamefont {Radzihovsky}, \citenamefont
  {Merali},\ and\ \citenamefont {Melko}}]{czischek2022dataenhanced}%
  \BibitemOpen
  \bibfield  {author} {\bibinfo {author} {\bibfnamefont {S.}~\bibnamefont
  {Czischek}}, \bibinfo {author} {\bibfnamefont {M.~S.}\ \bibnamefont {Moss}},
  \bibinfo {author} {\bibfnamefont {M.}~\bibnamefont {Radzihovsky}}, \bibinfo
  {author} {\bibfnamefont {E.}~\bibnamefont {Merali}},\ and\ \bibinfo {author}
  {\bibfnamefont {R.~G.}\ \bibnamefont {Melko}},\ }\bibfield  {title} {\bibinfo
  {title} {Data-enhanced variational {{Monte Carlo}} for {{Rydberg}} atom
  arrays},\ }\href@noop {} {\bibfield  {journal} {\bibinfo  {journal} {arXiv}\
  } (\bibinfo {year} {2022})},\ \Eprint {https://arxiv.org/abs/2203.04988}
  {arXiv:2203.04988} \BibitemShut {NoStop}%
\bibitem [{\citenamefont {Wecker}\ \emph {et~al.}(2015)\citenamefont {Wecker},
  \citenamefont {Hastings},\ and\ \citenamefont {Troyer}}]{wecker2015progress}%
  \BibitemOpen
  \bibfield  {author} {\bibinfo {author} {\bibfnamefont {D.}~\bibnamefont
  {Wecker}}, \bibinfo {author} {\bibfnamefont {M.~B.}\ \bibnamefont
  {Hastings}},\ and\ \bibinfo {author} {\bibfnamefont {M.}~\bibnamefont
  {Troyer}},\ }\bibfield  {title} {\bibinfo {title} {Progress towards practical
  quantum variational algorithms},\ }\href
  {https://doi.org/10.1103/PhysRevA.92.042303} {\bibfield  {journal} {\bibinfo
  {journal} {Phys. Rev. A}\ }\textbf {\bibinfo {volume} {92}},\ \bibinfo
  {pages} {042303} (\bibinfo {year} {2015})}\BibitemShut {NoStop}%
\bibitem [{\citenamefont {Gonthier}\ \emph {et~al.}(2020)\citenamefont
  {Gonthier}, \citenamefont {Radin}, \citenamefont {Buda}, \citenamefont
  {Doskocil}, \citenamefont {Abuan},\ and\ \citenamefont
  {Romero}}]{gonthier2020identifying}%
  \BibitemOpen
  \bibfield  {author} {\bibinfo {author} {\bibfnamefont {J.~F.}\ \bibnamefont
  {Gonthier}}, \bibinfo {author} {\bibfnamefont {M.~D.}\ \bibnamefont {Radin}},
  \bibinfo {author} {\bibfnamefont {C.}~\bibnamefont {Buda}}, \bibinfo {author}
  {\bibfnamefont {E.~J.}\ \bibnamefont {Doskocil}}, \bibinfo {author}
  {\bibfnamefont {C.~M.}\ \bibnamefont {Abuan}},\ and\ \bibinfo {author}
  {\bibfnamefont {J.}~\bibnamefont {Romero}},\ }\bibfield  {title} {\bibinfo
  {title} {Identifying challenges towards practical quantum advantage through
  resource estimation: {{The}} measurement roadblock in the variational quantum
  eigensolver},\ }\href@noop {} {\bibfield  {journal} {\bibinfo  {journal}
  {arXiv}\ } (\bibinfo {year} {2020})},\ \Eprint
  {https://arxiv.org/abs/2012.04001} {arXiv:2012.04001} \BibitemShut {NoStop}%
\bibitem [{\citenamefont {Kingma}\ and\ \citenamefont
  {Ba}(2017)}]{kingma2017adam}%
  \BibitemOpen
  \bibfield  {author} {\bibinfo {author} {\bibfnamefont {D.~P.}\ \bibnamefont
  {Kingma}}\ and\ \bibinfo {author} {\bibfnamefont {J.}~\bibnamefont {Ba}},\
  }\bibfield  {title} {\bibinfo {title} {Adam: A method for stochastic
  optimization},\ }\href@noop {} {\bibfield  {journal} {\bibinfo  {journal}
  {arXiv}\ } (\bibinfo {year} {2017})},\ \Eprint
  {https://arxiv.org/abs/1412.6980} {arXiv:1412.6980} \BibitemShut {NoStop}%
\bibitem [{\citenamefont {Innes}(2018)}]{innes2018flux}%
  \BibitemOpen
  \bibfield  {author} {\bibinfo {author} {\bibfnamefont {M.}~\bibnamefont
  {Innes}},\ }\bibfield  {title} {\bibinfo {title} {Flux: {{Elegant}} machine
  learning with {{Julia}}},\ }\href {https://doi.org/10.21105/joss.00602}
  {\bibfield  {journal} {\bibinfo  {journal} {J. Open Source Softw.}\ }\textbf
  {\bibinfo {volume} {3}},\ \bibinfo {pages} {602} (\bibinfo {year}
  {2018})}\BibitemShut {NoStop}%
\bibitem [{\citenamefont {Innes}\ \emph {et~al.}(2018)\citenamefont {Innes},
  \citenamefont {Saba}, \citenamefont {Fischer}, \citenamefont {Gandhi},
  \citenamefont {Rudilosso}, \citenamefont {Joy}, \citenamefont {Karmali},
  \citenamefont {Pal},\ and\ \citenamefont {Shah}}]{innes2018fashionable}%
  \BibitemOpen
  \bibfield  {author} {\bibinfo {author} {\bibfnamefont {M.}~\bibnamefont
  {Innes}}, \bibinfo {author} {\bibfnamefont {E.}~\bibnamefont {Saba}},
  \bibinfo {author} {\bibfnamefont {K.}~\bibnamefont {Fischer}}, \bibinfo
  {author} {\bibfnamefont {D.}~\bibnamefont {Gandhi}}, \bibinfo {author}
  {\bibfnamefont {M.~C.}\ \bibnamefont {Rudilosso}}, \bibinfo {author}
  {\bibfnamefont {N.~M.}\ \bibnamefont {Joy}}, \bibinfo {author} {\bibfnamefont
  {T.}~\bibnamefont {Karmali}}, \bibinfo {author} {\bibfnamefont
  {A.}~\bibnamefont {Pal}},\ and\ \bibinfo {author} {\bibfnamefont
  {V.}~\bibnamefont {Shah}},\ }\bibfield  {title} {\bibinfo {title}
  {Fashionable modelling with {{Flux}}},\ }\href@noop {} {\bibfield  {journal}
  {\bibinfo  {journal} {arXiv}\ } (\bibinfo {year} {2018})},\ \Eprint
  {https://arxiv.org/abs/1811.01457} {arXiv:1811.01457} \BibitemShut {NoStop}%
\bibitem [{\citenamefont {Smith}\ \emph {et~al.}(2020)\citenamefont {Smith},
  \citenamefont {Burns}, \citenamefont {Simmonett}, \citenamefont {Parrish},
  \citenamefont {Schieber}, \citenamefont {Galvelis}, \citenamefont {Kraus},
  \citenamefont {Kruse}, \citenamefont {Di~Remigio}, \citenamefont {Alenaizan},
  \citenamefont {James}, \citenamefont {Lehtola}, \citenamefont {Misiewicz},
  \citenamefont {Scheurer}, \citenamefont {Shaw} \emph
  {et~al.}}]{smith2020psi4}%
  \BibitemOpen
  \bibfield  {author} {\bibinfo {author} {\bibfnamefont {D.~G.~A.}\
  \bibnamefont {Smith}}, \bibinfo {author} {\bibfnamefont {L.~A.}\ \bibnamefont
  {Burns}}, \bibinfo {author} {\bibfnamefont {A.~C.}\ \bibnamefont
  {Simmonett}}, \bibinfo {author} {\bibfnamefont {R.~M.}\ \bibnamefont
  {Parrish}}, \bibinfo {author} {\bibfnamefont {M.~C.}\ \bibnamefont
  {Schieber}}, \bibinfo {author} {\bibfnamefont {R.}~\bibnamefont {Galvelis}},
  \bibinfo {author} {\bibfnamefont {P.}~\bibnamefont {Kraus}}, \bibinfo
  {author} {\bibfnamefont {H.}~\bibnamefont {Kruse}}, \bibinfo {author}
  {\bibfnamefont {R.}~\bibnamefont {Di~Remigio}}, \bibinfo {author}
  {\bibfnamefont {A.}~\bibnamefont {Alenaizan}}, \bibinfo {author}
  {\bibfnamefont {A.~M.}\ \bibnamefont {James}}, \bibinfo {author}
  {\bibfnamefont {S.}~\bibnamefont {Lehtola}}, \bibinfo {author} {\bibfnamefont
  {J.~P.}\ \bibnamefont {Misiewicz}}, \bibinfo {author} {\bibfnamefont
  {M.}~\bibnamefont {Scheurer}}, \bibinfo {author} {\bibfnamefont {R.~A.}\
  \bibnamefont {Shaw}}, \emph {et~al.},\ }\bibfield  {title} {\bibinfo {title}
  {Psi4 1.4: {{Open-source}} software for high-throughput quantum chemistry},\
  }\href {https://doi.org/10.1063/5.0006002} {\bibfield  {journal} {\bibinfo
  {journal} {J. Chem. Phys.}\ }\textbf {\bibinfo {volume} {152}},\ \bibinfo
  {pages} {184108} (\bibinfo {year} {2020})}\BibitemShut {NoStop}%
\bibitem [{\citenamefont {Hansen}\ and\ \citenamefont
  {Larsen}(1996)}]{hansen1996unsupervised}%
  \BibitemOpen
  \bibfield  {author} {\bibinfo {author} {\bibfnamefont {L.}~\bibnamefont
  {Hansen}}\ and\ \bibinfo {author} {\bibfnamefont {J.}~\bibnamefont
  {Larsen}},\ }\bibfield  {title} {\bibinfo {title} {Unsupervised learning and
  generalization},\ }in\ \href {https://doi.org/10.1109/ICNN.1996.548861}
  {\emph {\bibinfo {booktitle} {Proceedings of {{IEEE International
  Conference}} on {{Neural Networks}} ({{ICNN}}'96)}}},\ Vol.~\bibinfo {volume}
  {1}\ (\bibinfo  {publisher} {{IEEE}},\ \bibinfo {address} {{Washington, DC,
  USA}},\ \bibinfo {year} {1996})\ pp.\ \bibinfo {pages} {25--30}\BibitemShut
  {NoStop}%
\end{thebibliography}
\end{document}